\documentclass[aps,pre,amsmath,amsfonts,amssymb,twocolumn]{revtex4}
\usepackage{bm}        

\usepackage[latin1]{inputenc}
\usepackage{graphicx}
\usepackage[english]{babel}
\usepackage[usenames,dvipsnames]{color}

\usepackage{amsfonts}
\usepackage{amsmath}
\usepackage{amssymb}

\usepackage{color}

\usepackage{soul}

\newcommand{\beq}{\begin{equation}}
\newcommand{\eeq}{\end{equation}}
\newcommand{\nn}{\nonumber}
\newcommand{\ket}[1]{|#1\rangle}
\newcommand{\bra}[1]{\langle #1|}

\newcommand{\ra}{\rightarrow}

\makeatletter
\@ifundefined{textcolor}{}
{%
 \definecolor{BLACK}{gray}{0}
 \definecolor{WHITE}{gray}{1}
 \definecolor{RED}{rgb}{1,0,0}
 \definecolor{GREEN}{rgb}{0,1,0}
 \definecolor{BLUE}{rgb}{0,0,1}
 \definecolor{CYAN}{cmyk}{1,0,0,0}
 \definecolor{MAGENTA}{cmyk}{0,1,0,0}
 \definecolor{YELLOW}{cmyk}{0,0,1,0}
 }

\begin{document}


\title{Quantum dissipative adaptation}

\author{Daniel Valente
$^{1}$
}
\email{valente.daniel@gmail.com}

\author{Frederico Brito
$^{2}$
}

\author{Thiago Werlang
$^{1}$
}

\affiliation{
$^{1}$ 
Instituto de F\'isica, Universidade Federal de Mato Grosso, Cuiab\'a 78060-900 Mato Grosso, Brazil
}

\affiliation{
$^{2}$ 
Instituto de F\'isica de S\~ao Carlos, Universidade de S\~ao Paulo, Caixa Postal 369, S\~ao Carlos 13560-970 S\~ao Paulo, Brazil
}


\maketitle
\section{Abstract}
Dissipative adaptation is a general thermodynamic mechanism that explains self-organization in a broad class of driven classical many-body systems.
It establishes how the most likely (adapted) states of a system subjected to a given drive tend to be those following trajectories of highest work absorption, followed by dissipated heat to the reservoir.
Here, we extend the dissipative adaptation phenomenon to the quantum realm.
We employ a fully-quantized exactly solvable model, where the source of work on a three-level system is a single-photon pulse added to a zero-temperature infinite environment, a scenario that cannot be treated by the classical framework.
We find a set of equalities relating adaptation likelihood, absorbed work, heat dissipation and variation of the informational entropy of the environment.
Our proof of principle provides the starting point towards a quantum thermodynamics of driven self-organization.

\section{Introduction}
When a physical system is simultaneously subjected to both predictable and random energy exchanges, what dictates the likelihood of a given state to be found?
From a classical thermodynamic perspective, where energy exchanges are classified as work or heat, the concept of  dissipative adaptation has recently been put forward by J. England as the expected answer, at least in the context of driven self-assembly \cite{naturenano2015}.
Qualitatively speaking, dissipative adaptation establishes that, given a certain drive (an external work source), the most adapted (most likely and lasting) states of a fluctuating system tend to be those with a history of exceptionally high work absorption followed by heat dissipation to the environment.
Because heat dissipation is an irreversible process, the higher the dissipation, the less likely the reverse trajectory is.
In the long run, the system may appear to us as self-organized in this drive-dependent state of highest energy-consuming history.
Dissipative adaptation is the recent theoretical development of a long search for the emergence of order from disorder, as inspired by life-like behaviour \cite{prigogine,JCP2013}.
Examples revealing this general mechanism of energy-consuming irreversible self-organization span diverse systems, environments, lengths and timescales, as shown both theoretically \cite{PRL2017,PNAS2017,ragazzon18} and experimentally \cite{science2010, scirep2013, PRE2015, naturephoton2018, naturenano2020, naturephys2020}.

The dissipative adaptation phenomenon has been originally formulated in terms of fluctuation theorems.
Fluctuation theorems are equalities relating out-of-equilibrium processes with thermal-equilibrium variables, giving evidence that the fluctuations present in many realizations of a process can provide useful knowledge when summed up.
The pioneering example is the so called Jarzynski equality \cite{jarzynski}, $\langle \exp(-\beta W_{\mathrm{abs}}) \rangle = \exp(-\beta \Delta F)$, where $\beta=1/(k_B T)$ is the inverse temperature, $W_{\mathrm{abs}}$ is the work absorbed by the system as described by a time-dependent Hamiltonian, $\Delta F$ is the variation in the Helmholtz free energy and the brackets is the ensemble average over realizations of the process, initially departing from thermal equilibrium.
The Jarzynski equality has been lately derived by Crooks from what he called a microscopically reversible condition \cite{crooks},

\beq
\frac{ p_{i\ra j}(t) }{ p^*_{j \ra i}(\tau-t) } = e^{\beta Q_{\mathrm{diss}}},
\label{mrc}
\eeq
where the forward $p_{i\ra j}(t)$ and backward $p^*_{j\ra i}(\tau-t)$ probabilities for the system to follow paths linking states $i$ and $j$ are related with the amount of heat stochastically dissipated to the environment, $Q_{\mathrm{diss}}$ (a functional of the phase-space trajectory and of the driving protocol performing work on the system). $p^*_{j\ra i}(\tau-t)$ is computed with the reversed time-dependent protocol.
$\beta Q_{\mathrm{diss}}$ relates the statistical irreversibility with the thermodynamic entropy production.
Let us call $E_{ij} = E_j-E_i$ the energy difference between the specific final and initial states.
Energy conservation during each stochastic realization,

\beq
W_{\mathrm{abs}}=Q_{\mathrm{diss}}+ E_{ij},
\eeq
gives the hint on how the work source fuels the dissipative adaptation.
The higher the absorbed work, the more heat can be released, hence the more irreversible the path can become.
To emphasize this adaptation-work relation, England rearranges Eq. (\ref{mrc}) as \cite{naturenano2015}

\beq
\frac{p_{i\ra j}(t) }{p_{i\ra k}(t) } = 
e^{-\beta E_{kj}} 
\frac{p^*_{j\ra i}(\tau-t) }{p^*_{k\ra i}(\tau-t) } 
\frac{ 
\langle e^{-\beta W_{\mathrm{abs}}  } \rangle_{ik}
}{
\langle e^{-\beta W_{\mathrm{abs}} }  \rangle_{ij} },
\label{cda}
\eeq
where the angle brackets denote a weighted average over all microtrajectories with fixed start ($i$) and end ($j,k$) points (the coarse graining over microtrajectories describes experimentally accessible states of out-of-equilibrium self-organizing classical many-body systems; see Perunov {\it et al.} \cite{prx2016} for further details).
Equation (\ref{cda}) establishes the classical theoretical framework behind dissipative adaptation by showing how a given final state can be statistically privileged by work consumption.

Here, we extend the concept of dissipative adaptation to the quantum realm.
Our main goal is to test the robustness of the dissipative adaptation concept beyond its original theoretical framework discussed above.
From a technical viewpoint, at vanishing temperatures ($\beta \ra \infty$), where quantum fluctuations and correlations usually prevail, Eqs. (\ref{mrc}) and (\ref{cda}) are ill defined.
We employ a system-plus-reservoir approach to derive the exact equations of motion of a three-level lambda ($\Lambda$) system driven by a single-photon pulse added to a zero-temperature environment.
We find that the most adapted (self-organized) quantum state of the lambda system is indeed that with a history of highest work absorption followed by maximal heat dissipation, thus characterizing a dissipative adaptation.
As a consequence of our work, we establish the starting point of a quantum thermodynamics of driven self-organization, so far unexplored to the best of our knowledge.
We hope that the notion of a quantum dissipative adaptation may also provide fresh insights to quantum biology \cite{lambert}, not only because adaptation and self-organization are concepts inspired in life-like behaviour, but also because our results may find applications to discussions on quantum signatures in photosynthesis \cite{AOC14, naturephys2014,naturecomm2018wallraff, scirep2019}.

\section{Results}

\begin{figure}[!htb]
\centering
\includegraphics[width=1.0\linewidth]{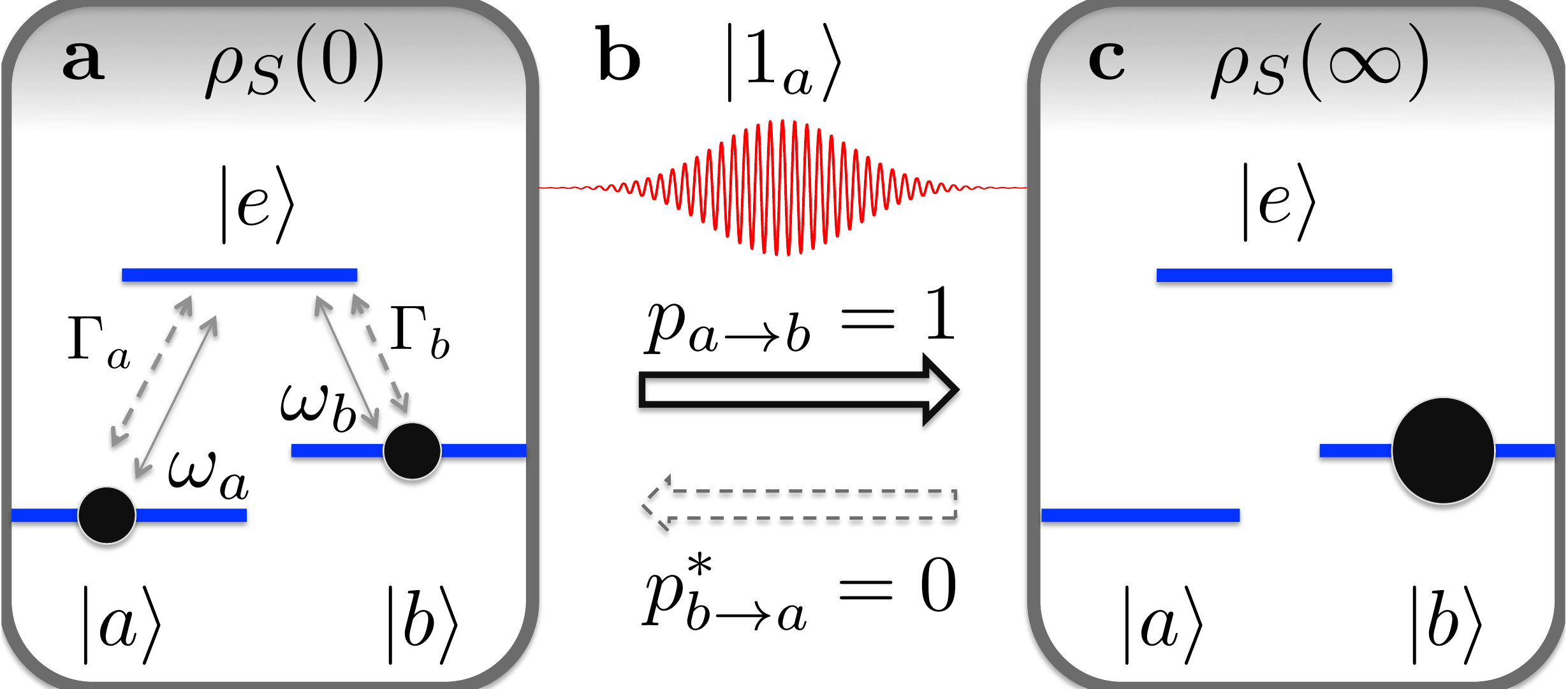} 
\caption{
{\bf Quantum dissipative adaptation in a driven self-organized quantum state.}
{\bf a} Three-level system in lambda configuration is described by the density operator $\rho_S(t)$ at time $t$, starting at $t=0$. 
The energy eigenstates are $\ket{a}$, $\ket{b}$ and $\ket{e}$ (blue horizontal bars). 
The transition frequencies are $\omega_{a,b}$ (full gray arrows).
The environment-induced spontaneous emission rates are $\Gamma_{a,b}$ (dashed gray arrows).
The initial state is a mixture between $\ket{a}$ and $\ket{b}$, 
$\rho_S(0) = p_a^{(0)}\ket{a}\bra{a} + p_b^{(0)}\ket{b}\bra{b}$ (smaller black dots).
{\bf b} A single-photon pulse $\ket{1_a}$ (the work source) drives the lambda system dynamics, inducing the time-dependent transition probability $p_{a \ra b}(t)$ from $\ket{a}$ to $\ket{b}$ (the full black horizontal arrow represents the forward dynamics from time $t=0$ to $t\ra \infty$).
The backward probability (with a time-reversed pulse), $p^*_{b \ra a}(t) = 0$, vanishes at zero temperature (the dashed gray horizontal arrow represents the prohibited time-reversed transition).
{\bf c} The driven lambda system undergoes an ideal irreversible self-organizing dynamics ($p_{a \ra b}(\infty) = 1$ and $p^*_{b \ra a}(t)=0$, in panel {\bf b}), so that the asymptotic state is pure, $\rho_S(\infty) = \ket{b} \bra{b}$ (larger black dot), conditioned to maximizing the work absorbed and the heat dissipated in the $\ket{a}$ to $\ket{b}$ transition.
}
\label{fig1}
\end{figure}

{\bf Self-organized quantum state.}
To achieve our main goal, we look for the most elementary scenario where the quantum state of a certain physical system irreversibly self-organizes, as induced by the absorption of energy from an external drive, the excess of which is dissipated to the environment.
We choose a three-level system in $\Lambda$ configuration, labeled as $\ket{a}$, $\ket{b}$ and $\ket{e}$ (being $\ket{e}$ the most excited state, respectively with transition frequencies $\omega_{a,b}$) (see Fig.\ref{fig1}).
To keep the model exactly solvable for both the system and the environment, we assume that the drive source is a single-photon pulse added to the vacuum state of an infinite environment at zero temperature, $T=0$.
The environment induces spontaneous emission rates, $\Gamma_a$ and $\Gamma_b$.
Our main results are that

\begin{eqnarray}
p_{a\ra b}(\infty) &=& \frac{\Gamma_b}{\Gamma_a +\Gamma_b} \ \frac{\langle W_{\mathrm{abs}} \rangle_{a}}{\hbar \omega_a}, \ \mbox{and} \nn \\
p^*_{b\ra a}(t) &=& 0,
\label{mainresults}
\end{eqnarray}
showing that adaptation likelihood at long times, $p_{a\ra b}(t\ra\infty)$, 
is linearly proportional to the average absorbed work $\langle W_{\mathrm{abs}} \rangle_{a}$ from a single-photon pulse of arbitrary shape, resonant with $\omega_a$.
We call attention to the fact that this result is not immediately expected, since 
(i) the excitation probability is minimized ($p_{a \ra e}(t) \ll 1$, at all times) when the work absorption is maximized ($p_{a\ra b}(\infty) \ra 1$), so the process cannot be thought of as a simple absorption-plus-emission picture, and 
(ii) the final self-organized state is not restrained to be the ground state, which amounts to say that $\langle W_{\mathrm{abs}} \rangle_{a}$ does not depend on 
$\omega_b$, so the work is not related to $\hbar (\omega_a-\omega_b)$, in particular.
The second line in Eqs.(\ref{mainresults}) characterizes the irreversibility of the process.
The absorbed work is partly dissipated to the environment in the form of heat, $\langle Q_{\mathrm{diss}} \rangle_{a}$, satisfying energy conservation

\beq
\langle W_{\mathrm{abs}} \rangle_{a} = \langle Q_{\mathrm{diss}} \rangle_{a}
+ \langle H_{S}(\infty) \rangle - \langle H_{S}(0) \rangle.
\label{ec1}
\eeq
$H_S$ here is the Hamiltonian of the system.
Finally, we find an exact expression for 
the informational entropy of the environment at long times, $S_E(\infty)$, as a function of the average dissipated heat, $\langle Q_{\mathrm{diss}} \rangle_{a}$.
The entropy analysis here provides us with both a clearer physical picture of the process and an additional signature of the dissipative adaptation.

Let us suppose that our generic three-level system in $\Lambda$ configuration, with lowest-energy states $\ket{a}$ and $\ket{b}$ and excited state $\ket{e}$, is initially in a non-driven steady state, in contact with the environment at temperature $T=0$.
In this case, at precisely $T=0$, the steady state is not uniquely defined, even for non-degenerated cases.
We can choose its initial quantum state to be a mixture of the lowest-energy states, as described by the density operator

\beq
\rho_S(0) = 
p_a^{(0)} \ket{a}\bra{a} +
p_b^{(0)} \ket{b}\bra{b},
\label{rhozero}
\eeq
where $p_{a,b}^{(0)}$ depend on the preparation scheme.
To give a concrete example, in the preparation by means of a spontaneous emission starting at $\ket{e}$, one has $p_{a,b}^{(0)} = \Gamma_{a,b}/(\Gamma_a+\Gamma_b)$.
Now we look for the most elementary out-of-equilibrium stochastic process that drives the system from this (generally mixed) initial state into a final (ideally pure) target state, let us say into state $\ket{b}$,

\beq
\rho_S(0) \ra \rho_S(\infty) = \ket{b}\bra{b}.
\label{selforg}
\eeq
In order to guarantee that the process is irreversible, in the light of Crooks condition, we should also apply the time-reversed  drive on the system departing from state $\rho_S(\infty) = \ket{b}\bra{b}$ and find that $\rho^*_S(\infty - t) \neq \rho_S(0)$ for $t \ra \infty$.

We employ a system-plus-reservoir approach, where we assume a global time-independent Hamiltonian of the system and its environment,

\beq
H=H_S + H_I + H_E.
\label{global}
\eeq
As we show in what follows, we find a self-organized quantum state in the well-known dipolar model of light-matter interaction in the rotating-wave approximation \cite{dv2012,chen2004},

\beq
H_I = -i\hbar \sum_\omega (g_a a_\omega \sigma_a^\dagger + g_b b_\omega \sigma_b^\dagger - \mbox{H.c.}).
\label{HI}
\eeq
Here, $\sigma_k = \ket{k}\bra{e}$ (for $k=a,b$) and $\mbox{H.c.}$ is the Hermitian conjugate. 
We consider a continuum of frequencies, $\sum_\omega \ra \int d\omega \varrho_\omega$, with density of modes $\varrho_\omega$.
Modes $\left\{ a_\omega \right\} $ and $\left\{ b_\omega \right\}$ are the quantized field modes respectively interacting with the transitions $\ket{a}$ to $\ket{e}$ and $\ket{b}$ to $\ket{e}$.
The continuum of frequencies allows us to employ a Wigner-Weisskopf approximation to obtain dissipation rates
$\Gamma_{a} = 2\pi g_a^2 \varrho_{\omega_{a}}$ and 
$\Gamma_{b} = 2\pi g_b^2 \varrho_{\omega_{b}}$.
Finally, 
$
H_E = \sum_\omega \hbar \omega (a_\omega^\dagger a_\omega + b_\omega^\dagger b_\omega)
$
and

\beq
H_S = \hbar \omega_a \ket{e}\bra{e} + \hbar \delta_{ab} \ket{b}\bra{b},
\eeq
where $\hbar\delta_{ab} = \hbar(\omega_a - \omega_b)$ is the energy difference between states $\ket{b}$ and $\ket{a}$.
It is worth emphasizing that our main results in this paper are independent of $\delta_{ab}$.

To keep the model exactly solvable for both the system and the environment, we choose the drive as provided by a 
propagating pulse containing a single photon.
We choose the photon to initially populate only the continuum of modes $\left\{ a_\omega \right\}$, so the vacuum state of $\left\{ b_\omega \right\}$ allows us to avoid depleting our target state $\ket{b}$.
The initial state of the field is

\beq
\ket{1_a} = \sum_\omega \phi_\omega^{a}(0) \ a_\omega^\dagger \ket{0}.
\label{1a}
\eeq
$\ket{0} = \prod_\omega \ket{0_\omega^{a}} \otimes \ket{0_\omega^{b}}$ is the vacuum state of all the field modes.
The initial state of the $\Lambda$ system is the mixed $\rho_S(0)$ given in Eq.(\ref{rhozero}).
The global quantum state is given by

\beq
\rho(t) = U\left( \rho_S(0) \otimes \ket{1_a} \bra{1_a} \right) U^\dagger,
\label{rhoglobal}
\eeq
where $U=\exp(-iH t/\hbar)$ for the (time-independent) global Hamiltonian $H$ [Eq.(\ref{global})].
The quantum states of the system and the environment are obtained by the partial traces 
$\rho_S(t) = \mbox{tr}_E[\rho(t)]$
and
$\rho_E(t) = \mbox{tr}_S[\rho(t)]$.

We obtain analytical expressions for the probabilities $p_k(t) = \bra{k} \rho_S(t) \ket{k}$, for $k=a,b,e$.
Equation (\ref{rhozero}) allows us to write the reduced state of the system as $\rho_S(t) = p_a^{(0)} \mbox{tr}_E \left[ U\ket{a, 1_a}\bra{a,1_a}U^\dagger \right] + p_b^{(0)} \mbox{tr}_E \left[ U\ket{b,1_a} \bra{b,1_a}U^\dagger \right]$.
It proves useful to write $p_k(t)$ in terms of transition probabilities, $p_{n\ra k}(t)$, as

\beq
p_k(t) = \sum_n p_n^{(0)} p_{n\ra k}(t),
\label{sop}
\eeq
where $n=a,b$,
and 
\beq
p_{n\ra k}(t) \equiv \bra{k} \mbox{tr}_E \left[ U\ket{n, 1_a}\bra{n, 1_a}U^\dagger \right] \ket{k}.
\label{pnk}
\eeq
Equations (\ref{sop}) and (\ref{pnk}), which follow directly from the mixed-state structure of the initial state, unravel a close analogy with the notation used in Eq.(\ref{mrc}) for the microscopically reversible condition.
Equation (\ref{sop}) has no relation with the degree of Markovianity in the dynamics of the reduced state $\rho_S(t)$, though.
On the contrary, Valente {\it et al.} \cite{nonmarkov} characterize non-Markovianity in a quite similar scenario.
The photon initially at modes $\left\{ a_\omega\right\}$ does not interact with the three-level system initially at $\ket{b}$ [as derived from Eqs.(\ref{rhozero}), (\ref{HI}) and (\ref{rhoglobal})], so we find that 
$p_{b\ra a}(t) = p_{b\ra e}(t) = 0$ and $p_{b\ra b}(t) = 1$.
Because the transition probabilities vanish identically, regardless of the initial pulse shape, the backwards probability also vanishes, 

\beq
p^*_{b\ra a}(t) = 0.
\label{rev}
\eeq
We have defined the reverse drive protocol here as the mirrored shape of the initial pulse (further discussed below).
Put simply, we revert only the drive, not the entire universe.
The final global state of the system plus the environment is obviously reversible in our model since it is governed by the global unitary operator $U$ in Eq.(\ref{rhoglobal}).
Equation (\ref{rev}) explains the second line in Eqs.(\ref{mainresults}).
We now compute 

\beq
p_{a\ra k}(t) = \bra{k} \mbox{tr}_E \left[ \ket{\xi(t)} \bra{\xi(t)} \right] \ket{k},
\eeq
where $\ket{\xi(t)} \equiv U \ket{a, 1_a}$.
Since our $H$ conserves the total number of excitations, we can restrict our model to the one-excitation subspace, 
$
\ket{\xi(t)} = \psi(t) \ket{e,0} 
+ \sum_\omega \phi_\omega^a(t) a_\omega^\dagger \ket{a,0}
+ \phi_\omega^b(t) b_\omega^\dagger \ket{b,0}.
$
We find that 

\begin{align}
p_{a\ra e}(t) &= |\psi(t)|^2, \nn \\
p_{a\ra b}(t) &= \sum_\omega |\phi^b_\omega(t)|^2 \nn \\
& = \frac{1}{2\pi \varrho c} \int_{-\infty}^{\infty} dz \ |\phi_{b}(z,t)|^2,
\label{pae}
\end{align}
and similarly with $p_{a\ra a}(t)$. 
Without loss of generality, we have defined a one-dimensional real-space representation for the amplitudes,
$\phi_n(z,t) \equiv \sum_\omega \phi_\omega^n(t) \exp(ik_\omega z)$, which characterizes the pulse shape.
We have also employed a linear dispersion relation, 
$\omega = c k_\omega$, 
and approximated the density of modes by a constant, 
$\varrho_\omega \approx \varrho$.
Note that, in Eq.(\ref{HI}), we have implicitly assumed the three-level system to be positioned at $z_S = 0$; otherwise, a phase term such as $\exp(i k_\omega z_S)$ should have been included in the sum over modes.

To go a step further, as to obtain explicit expressions for the transition probabilities, we solve the Schr\"odinger equation for 
$\ket{\xi(t)}$ (see Methods).
Our intention with keeping modes $\left\{b_\omega \right\}$ initially in the vacuum state, as we did in Eq.(\ref{1a}), was to minimize excitations promoting the unwanted backward ($\ket{b} \ra \ket{a}$) transitions.
Now that we have defined the amplitudes $\phi_n(z,t)$, we see that Eq.(\ref{1a}) implies $\phi_b(z,0) = 0$, which we combine with (\ref{phib}) and substitute in (\ref{pae}).
After a change of variables, we find that

\beq
p_{a\ra b}(t) = \Gamma_b \int_{0}^{t} dt' |\psi(t')|^2.
\label{pabt}
\eeq
Although the main results in this paper do not depend on the initial pulse shape in modes $\left\{ a_\omega\right\}$ (i.e., on the choice of $|\phi_a(z,0)|$), it is worth working out an explicit example.
To that end, we now set $\phi_a(z,0)$ to have an exponential envelope profile of linewidth $\Delta$ and a central frequency $\omega_L$ (see Eqs. (\ref{shape}) and (\ref{exp}) in the Methods), as typical in spontaneous emission.
We are particularly interested in the resonant condition $\omega_L = \omega_a$ (see the Methods for the more general solution).
We find, in the monochromatic limit, $\Delta \ll \Gamma_a + \Gamma_b$, and at long times $t \gg \Delta^{-1}$, that

\beq
p_{a\ra b}(\infty) = \Gamma_b \ \frac{4\Gamma_a}{(\Gamma_a + \Gamma_b)^2}.
\label{pabtinfty}
\eeq
When $\Gamma_a = \Gamma_b$, we have that $p_{a\ra b}(\infty) = 1$.
Equation (\ref{pabtinfty}) reveals the ideal driven self-organization we have been looking for.
As stated before, the self-organization in our model does not depend on $\delta_{ab}$.
As a final step, we combine Eq. (\ref{pabtinfty}) and $p_{b\ra b}(t) = 1$ to see that, in the ideal (monochromatic and resonant) regime,

\beq
p_b(\infty) = p_a^{(0)} p_{a\ra b}(\infty) + p_b^{(0)} p_{b\ra b}(\infty) = 1,
\eeq
that is, $\rho_S(\infty) = \ket{b}\bra{b}$ and $\rho^*_S(\infty-t) = \ket{b}\bra{b}$ for all times $t$. 
Hence, in the $t \ra\infty$ limit, $\rho^*_S(\infty-t) \neq \rho_S(0)$, as was to be shown.

{\bf Energetics of the self-organization.}
We now need to verify that the self-organized quantum state we have found can indeed be classified as dissipative adaptation. 
We shall find that ideal self-organization costs maximal work absorption, followed by maximal dissipation.
This is not an obvious relation because in the ideal self-organization (which takes place in the monochromatic limit, as we have shown above), the excitation probability is minimized (rather than maximized), 

\beq
p_{a\ra e}(t) = |\psi(t)|^2 \ll 1.
\eeq
For instance, in the case of an incoming pulse with exponential profile (as used in Eq.(\ref{pabtinfty})), $|\psi(t)|^2 \leq 4 \Delta \Gamma_a/(\Gamma_a+\Gamma_b)^2 \ll 1$, for $\Delta \ll \Gamma_a + \Gamma_b$.
Extremely low excitation probabilities may leave the false impression that no energy is absorbed neither dissipated at all.
To address this issue, we need to resolve energy transfer into work and heat in our model.
As mentioned earlier, work and heat can be regarded as the predictable versus the random energy exchanges, respectively.
We restrict ourselves once again to the resonant case, $\omega_L=\omega_a$, keeping in mind however an arbitrary pulse shape.
The average energy of the $\Lambda$ system driven by this resonant photon is given by

\beq
\langle H_S(t) \rangle \equiv \mbox{tr}[\rho_S(t) H_S]  = p_e(t) \hbar\omega_a + p_b(t) \hbar \delta_{ab}.
\label{averagehs}
\eeq
The resonant condition here avoids dynamic Stark shifts that would otherwise bring extra (dispersive) energetic contributions from time-dependent frequencies, as shown by Valente {\it et al.} \cite{OL2018}.
To be more precise, dispersive (refractive) energetic contributions depend on the average interaction energy \cite{OL2018}, which vanishes at resonance ($\omega_L = \omega_a$ implies that $\langle H_I(t) \rangle \equiv \mbox{tr}[\rho(t) H_I] = 0$; see details in the Methods).
That justifies why we have used only $H_S$ in Eq.(\ref{averagehs}).
As inspired by Eq.(\ref{cda}), we are interested in the work performed on the system during the dynamical transition starting from $\ket{a}$ and arriving at $\ket{b}$.
Therefore, we now consider $\rho_S(0) = \ket{a} \bra{a}$, for which $p_e(t) = |\psi(t)|^2$.
From the Schr\"odinger equation (see Eq.(\ref{eqpsi}) in the Methods), we find that

\beq
\partial_t p_e = - (\Gamma_a + \Gamma_b) p_e - 2 g_a  \mbox{Re} \left[ \phi_a(-ct,0) \psi^*(t) \right],
\label{exc}
\eeq
where $\mbox{Re}\left[ \bullet \right]$ stands for the real part.
Equation (\ref{exc}) clearly shows us that the excited state of the $\Lambda$ system is governed by a (predictable) drive-dependent term, related to $\phi_a(-ct,0)$, and a (random) spontaneous emission term, proportional to $\Gamma_a + \Gamma_b$.
This motivates us to define the total average absorbed work $\langle W_{\mathrm{abs}} \rangle_{a}$ and the total average absorbed heat $\langle Q_{\mathrm{abs}} \rangle_{a}$ in the dynamics starting at $\rho_S(0) = \ket{a} \bra{a}$ as

\beq
\langle H_S(\infty) \rangle - \langle H_S(0) \rangle = \langle {Q}_{\mathrm{abs}} \rangle_{a} + \langle {W}_{\mathrm{abs}} \rangle_{a},
\label{ec2}
\eeq
where

\beq
\langle W_{\mathrm{abs}} \rangle_{a} \equiv \hbar\omega_a \int_0^\infty  - 2 g_a \mbox{Re}[\phi_a(-ct,0) \psi^*(t)] dt
\label{W}
\eeq
and

\beq
\langle {Q}_{\mathrm{abs}} \rangle_{a} \equiv \int_0^\infty -(\Gamma_a + \Gamma_b)p_e(t)\hbar\omega_a dt + \dot{p}_b \hbar\delta_{ab}dt. 
\label{Q}
\eeq
Our definition of work, Eq. (\ref{W}), can also be written as $\langle W_{\mathrm{abs}} \rangle_{a} = \int_0^\infty \langle \left(\partial_t{d}(t)\right) E_{\mathrm{in}}(t) \rangle dt$ (in the Heisenberg picture and in the rotating-wave approximation), revealing a more clear link to our classical notion of work.
Here, $d(t) = U^\dagger d\ U$ is the dipole operator, where $d=\sum_k d_{ek}\sigma_k+ \mbox{H.c.}$, the driving field operator at the system's position is $E_{\mathrm{in}}(t) = \sum_{\omega}i\epsilon_\omega\left(a_\omega+b_\omega\right)\exp{(-i\omega t)} + \mbox{H.c.}$, the coupling is $g_a = d_{ea}\epsilon_{\omega_a}/\hbar$, and the average is calculated at the initial state $\ket{a,1_a}$.
The Heisenberg picture also explains why work can be finite even though the average driving field is precisely zero at the single-photon state (i.e., $\bra{1_a} E_{\mathrm{in}}(t) \ket{1_a} = 0$).
This shows how the dissipative adaptation can be robust to a phase-incoherent work source, in contrast to the classical forces of well-defined phases used in \cite{PRL2017} (in our model, a semiclassical driving field with a well-defined phase would correspond to an initial coherent, or Glauber, state $\ket{\alpha} = \prod_\omega \ket{\alpha_\omega}$, where $a_\omega\ket{\alpha_\omega} = \alpha_\omega \ket{\alpha_\omega}$).
In Eq.(\ref{Q}), $\Gamma_{a,b} \propto g^2_{a,b}$ reveals that the heat is related with the variance of the interaction energy, explaining why the vanishing average interaction energy does not hinder energy exchange in the form of heat and confirming its stochastic nature.
The energy conservation in Eq.(\ref{ec1}) results, of course, from defining 
$\langle {Q}_{\mathrm{abs}} \rangle_{a} \equiv - \langle {Q}_{\mathrm{diss}} \rangle_{a}$ in Eq.(\ref{ec2}).
Finally, $\langle H_S(0) \rangle = 0$ and $\langle H_S(\infty) \rangle = p_{a \ra b}(\infty) \hbar\delta_{ab}$.
We remind that, during the dynamical transition from $\ket{a}$ to $\ket{b}$, we have $p_e(t) = |\psi(t)|^2$, allowing us to establish an exact adaptation-energy relation between Eqs.(\ref{pabt}) and (\ref{Q}).
We finally find our quantum adaptation-work relation,

\beq
p_{a\ra b}(\infty) = \frac{\Gamma_b}{\Gamma_a +\Gamma_b} \ \frac{\langle W_{\mathrm{abs}} \rangle_{a}}{\hbar \omega_a},
\label{adaptationwork}
\eeq
valid for a photon of arbitrary pulse shape and resonant with $\omega_a$, as well as for arbitrary $\delta_{ab}$, as stated in Eqs.(\ref{mainresults}).
In the case of the exponential pulse used in Eq.(\ref{pabtinfty}), for instance, we find that
$\langle W_{\mathrm{abs}} \rangle_{a} = \hbar \omega_a \ 4\Gamma_a/(\Gamma_a+\Gamma_b+\Delta)$, maximized in the monochromatic regime ($\Delta \ll \Gamma_a+\Gamma_b$).
If in addition $\Gamma_a = \Gamma_b$, we get $\langle W_{\mathrm{abs}} \rangle_{a} = 2 \hbar \omega_a$, which is twice the initial average energy contained in the single-photon pulse.
This counterintuitive result reinforces the notion that work is the amount of energy transferred during a process, rather than the average energy stored in a system at a given time (after all, unitary evolutions with time-independent Hamiltonians conserve energy; here, $\partial_t \langle H(t) \rangle = 0$ implies that $\langle H_S(t) \rangle + \langle H_I(t) \rangle + \langle H_E(t) \rangle = p_b(0)\hbar\delta_{ab} + \hbar \omega_L$, so at resonance and for $p_b(0)=0$ we find $\langle H_S(t) \rangle \leq \hbar \omega_L$).
Equation (\ref{adaptationwork}) is our key signature of a quantum dissipative adaptation.
The seeming low-excitation issue ($|\psi(t)|^2 \ll 1$, in the monochromatic limit) has finally been clarified, given that the time integral of $|\psi(t)|^2$ (in Eq.(\ref{pabt})) is not only finite, but also linearly proportional to the work required for the quantum dissipative adaptation to take place.

{\bf Entropy in the self-organization.}
In the classical formulation of dissipative adaptation, entropy production plays a key role in establishing the connection between energy transfer and statistical irreversibility, requiring no detailed knowledge on the state of the environment.
Here, irreversibility is readily characterized by the asymmetry between the forward, $p_{a\ra b}(t)$, and the backward, $p^*_{b\ra a}(t)$, processes.
Nevertheless, we have the advantage that we keep the full description of the quantum state of the system plus the environment, $\rho(t)$ (at the expense of a greater degree of generality in our global Hamiltonian $H$).
With $\rho_E(t)=\mbox{tr}_S[\rho(t)]$ at hands (see Methods), we now seek to describe what happens to the environment during and after the drive interacts with the three-level system.
We calculate the exact expression for the von Neumann entropy, 

\beq
S_E(t) = - \mbox{tr}[\rho_E(t)\ln \rho_E(t)].
\eeq
See Eqs.(\ref{vNlambdas}) and (\ref{lambda34}) in the Methods for the analytic expression.
We have found that our classical intuition, namely, that better adaptation produces more entropy in the environment, can be found by an appropriate distinction between the classical and the quantum contributions to $S_E$.

The idea behind this distinction is the following.
Let us first suppose that the system is initially at $\ket{a}$.
A highly monochromatic incoming photon will fully induce transition from state $\ket{a}$ to $\ket{b}$.
Hence, the outgoing photon will be detected at modes $\left\{ b_\omega \right\}$, as well.
Now, by considering an initially mixed state of the three-level system as given by $(1/2)\ket{a}\bra{a} + (1/2)\ket{b}\bra{b}$, a highly monochromatic incoming photon will have $1/2$ probability of leaving at $\left\{ a_\omega \right\}$ (in the cases where it encounters the system at $\ket{b}$) and $1/2$ of leaving at $\left\{ b_\omega \right\}$ (in the cases where it encounters the system at $\ket{a}$), so the final global state would be mixed and separable,

\beq
\rho(\infty) \approx \ket{b}\bra{b} \otimes 
\left(
\frac{1}{2} \ket{\tilde{1}_b } \bra{\tilde{1}_b }
+
\frac{1}{2} \ket{1^{\mathrm{free}}_a } \bra{1^{\mathrm{free}}_a } 
\right).
\label{mixture}
\eeq
$\ket{\tilde{1}_b}$ and $\ket{1^{\mathrm{free}}_a }$ have been defined in Eqs.(\ref{tilde1k}) and (\ref{1afree}) below (see Methods).
This is what we call the classical contribution to the final mixed state of the field: the initial mixture of the system is fully transferred to the environment.

Let us now consider again that the system is initially at $\ket{a}$ (i.e., $p_a^{(0)}=1$).
However, let us assume that the linewidth of the pulse is of the order of the dissipation rates of the three-level system.
In that case, the final state of the global system becomes entangled,

\beq
\ket{\xi(\infty)} = {\sqrt{N_a}} \ket{a,\tilde{1}_a} + {\sqrt{N_b}} \ket{b,\tilde{1}_b},
\label{entang}
\eeq
$N_k$ being the average number of photons at modes $k$ (see Eq.(\ref{rhoE}) in the Methods).
Therefore, the quantum state of the environment in this case, $\rho_E(\infty) = \mbox{tr}_S[\ket{\xi(\infty)} \bra{\xi(\infty)}]$, is also mixed between modes $\left\{ a_\omega \right\}$ and $\left\{ b_\omega \right\}$. However, the mixture in this case arises from a sustained system-environment quantum entanglement rather than from the statistical mixture in the initial state of the system.

\begin{figure}[!htb]
\centering
\includegraphics[width=1.0\linewidth]{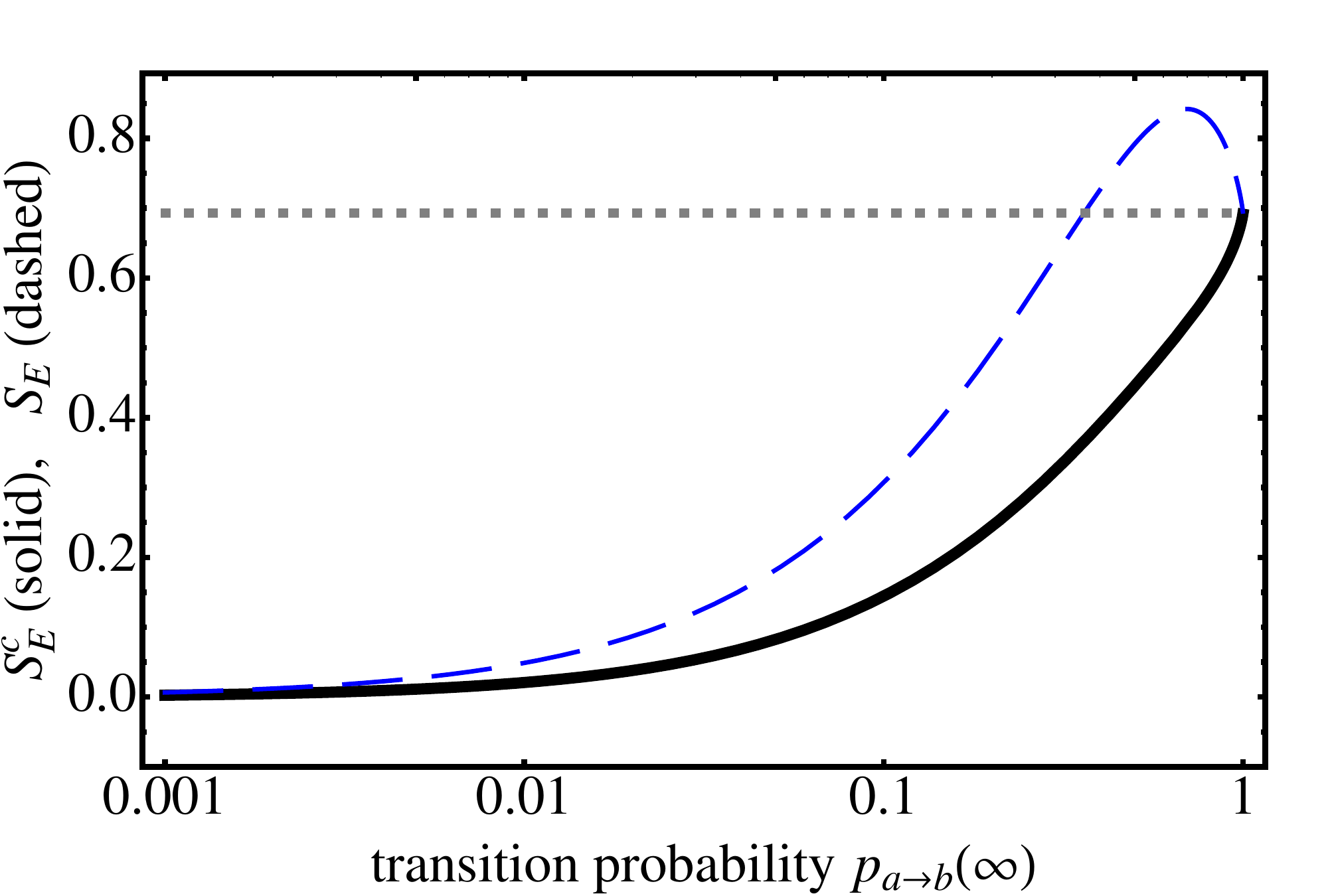} 
\caption{
{\bf Entropy of the environment.}
The classical contribution to the environment entropy, $S_E^c(\infty)$, increases monotonically as a function of the transition probability from $\ket{a}$ to $\ket{b}$, $p_{a \ra b}(\infty)$ (solid black), providing an additional signature of the dissipative adaptation.
The environment entropy, $S_E(\infty)$, presents a non-monotonic behaviour as a function of $p_{a \ra b}(\infty)$ (dashed blue), due to a sustained system-environment quantum entanglement that can contribute to the entropy production of the environment.
In both curves, we refer to the asymptotic state $t\ra \infty$ of the dynamics.
We use $\Gamma_a=\Gamma_b$ as the environment-induced spontaneous emission rates and $p_a^{(0)} = p_b^{(0)} = 1/2$ as the probabilities in the initial state of the lambda system, $\rho_S(0) = p_a^{(0)}\ket{a}\bra{a} + p_b^{(0)}\ket{b}\bra{b}$ (the curves do not qualitatively depend on these choices).
The dotted gray line is $\ln(2)$.
Note that both $S_E^c(\infty)$ as a function of $p_{a \ra b}(\infty)$ and $S_E(\infty)$ as a function of $p_{a \ra b}(\infty)$ do not explicitly depend on the photon pulse shape, though $p_{a \ra b}(t)$ does; e.g., attaining $p_{a \ra b}(\infty) \ra 1$ requires an extremely monochromatic (long) photon, whereas the $p_{a \ra b}(\infty) \ra 0$ limit is attained in the extremely broadband (short) pulse regime.
}
\label{fig2}
\end{figure}

To unravel these two entropy contributions, we define the classical contribution to the environment entropy as

\beq
S_E^c(t) \equiv S_E(t) - p_a^{(0)} S(\mbox{tr}_S[\ket{\xi(t)} \bra{\xi(t)}]),
\label{defsec}
\eeq
where 
$S(\bullet)\equiv -\mbox{tr}[\bullet \ln \bullet]$
is the von Neumann entropy (see Eq.(\ref{qvN}) in the Methods).
In Eq.(\ref{defsec}), we have taken into account that only the term proportional to $p_a^{(0)}$ in $\rho_S(0)$ generates the entanglement discussed in Eq.(\ref{entang}).
Now we focus on the long time limit, $t\ra\infty$.
Most interestingly, we have analytically expressed $S_E^c(\infty)$ as a function of $p_{a\ra b}(\infty)$ (see Eqs.(\ref{vNlambdas})-(\ref{overlap}) in the Methods). 
Figure (\ref{fig2}) illustrates that
$S_E^c(\infty)$ vs. $p_{a\ra b}(\infty)$ (solid black line) is a monotonic function, in contrast with the non-monotonic $S_E(\infty)$ vs. $p_{a\ra b}(\infty)$ (dashed blue line).
This monotonic behaviour shows that the more organized is the three-level system, the higher is the classical contribution to the entropy of the environment at long times $t\ra\infty$.
Besides having derived $S_E^c(\infty)$ as a function of $p_{a\ra b}(\infty)$, we have also expressed $p_{a\ra b}(\infty)$ as a function of the average dissipated heat in the $\ket{a}$ to $\ket{b}$ transition, $\langle {Q}_{\mathrm{diss}} \rangle_a$, with the help of Eq.(\ref{adaptationwork}).
This establishes the function $S_E(\infty)$ vs. $\langle {Q}_{\mathrm{diss}} \rangle_{a}$, as mentioned earlier.
Additionally, we have derived $S_E^c(\infty)$ as a function of $\langle {Q}_{\mathrm{diss}} \rangle_{a}$.
This is even more meaningful, because we have found that the function $S_E^c(\infty)$ vs. $\langle {Q}_{\mathrm{diss}} \rangle_{a}$ is monotonic.
To provide an example illustrating this monotonicity, let us take the degenerate case ($\delta_{ab}=0$) with equal dissipation rates ($\Gamma_a=\Gamma_b$).
In that case, we have that 
$p_{a\ra b}(\infty) = \langle {Q}_{\mathrm{diss}} \rangle_{a}/(2\hbar\omega_a)$, 
showing that $S_E^c(\infty)$ vs. $\langle {Q}_{\mathrm{diss}} \rangle_{a}$ is monotonic, since $S_E^c(\infty)$ vs. $p_{a\ra b}(\infty)$ is monotonic as well (Eq.(\ref{pQgen}) in the Methods provides the more general relation).
The monotonicity here strengthens the signature of dissipative adaptation.
Namely, maximal adaptation (irreversible self-organization) not only costs maximal work absorption (as we have shown in the energetics analysis), but also maximizes the dissipated heat which, in turn, maximizes the classical contribution to the environment entropy.


\section{Discussion}
Our results establish the quantum dissipative adaptation underlying the driven self-organization of a quantum state, going beyond the classical formulation.
We have explored an elementary fully-quantized model, exactly solvable for both the system and the environment, where the irreversibly self-organized quantum state of a three-level system is due to the work absorption from a single-photon pulse, part of which is dissipated to the environment (as shown in Eqs.(\ref{mainresults}) and (\ref{ec1})).
The irreversibility of this self-organization became clear from the asymmetry in the transition probabilities in the forward and backward processes.
Finally, with the purpose of providing an additional signature of dissipative adaptation, we have analytically investigated the environment entropy.
We have found that the classical contribution to the environment entropy variation is a monotonic function of $p_{a\ra b}(\infty)$ (as illustrated in Fig.\ref{fig2}), 
showing that maximizing adaptation not only requires maximal work absorption, but also leads to maximal increase in the classical contribution to the environment entropy, due to maximal dissipated heat.

We remind that the meaning of this increase in the environment entropy is that of a statistical mixing between the field modes $\left\{ a_\omega \right\}$ and $\left\{ b_\omega \right\}$, as we have analysed in Eq.(\ref{mixture}).
Our model's dynamics do not lead to the thermalization of the environment, which always remains out of equilibrium.
If we wish to take the analogy further, so as to give a thermodynamic meaning to the entropy increase, we will have to add extra ingredients.
In practice, there could be for instance a slight increase in the local temperature of the environment in the vicinity of the $\Lambda$ system.
To describe that kind of effect, the model should consider a finite heat capacity environment (in contrast to our infinite-size environment) and some auxiliary light-matter coupling mechanism that could effectively create an interaction between the many frequency modes of the light field.
Such features provide means for the environment state to eventually approximate a Gibbs state at some new finite temperature $T' > T = 0$ (producing even more entropy than that we have calculated from our model), in the spirit of Timpanaro {\it et al.} \cite{landi20}.

Before concluding, we would like to point out how our model can be significant to quantum many-body systems.
Our intention is to indicate how broadly applicable the concept of quantum dissipative adaptation may become.
Firstly, we notice that the $\Lambda$ structure of energy levels can arise from the quantization of collective degrees of freedom describing many interacting atoms and electrons, as happens in the so-called artificial atoms (e.g., electron-hole pairs in semiconductor quantum dots \cite{lodahl} and the quantized magnetic flux in superconducting rings \cite{nori}).
Lodahl {\it et al.} \cite{lodahl} and Gu {\it et al.} \cite{nori} also discuss how these artificial atoms can be driven by single-photon pulses propagating in one-dimensional waveguides, building the closest possible scenario to that in our model.

We can also envision the quantum dissipative adaptation in ensembles of non-interacting atoms or spins.
That is more easily seen when we realize that our idea of a driven self-organized quantum state is notably similar to the dynamics induced by stimulated Raman adiabatic passage (STIRAP) \cite{stirap2017}, a versatile and robust technique that has been performed, e.g., in ultracold gases, in doped crystals and in nitrogen-vacancy centers.
STIRAP consists of an efficient population transfer between two discrete quantum states of an ensemble of emitters (usually the lowest levels of $\Lambda$ systems) by coherently coupling them with two radiation fields (well-controlled classical pulses) through an unpopulated intermediate state.
The connection we have in mind between the driven self-organization provided by STIRAP and dissipative adaptation becomes more evident in light of the recent proposal for using STIRAP as a tool for spectral hole burning (SHB) in inhomogeneously broadened systems \cite{molmer2019}.
The reason is that the mechanism behind standard SHB \cite{shb} turns out to be precisely that of classical dissipative adaptation, as discussed by Kedia {\it et al.} \cite{bistableDA}.
Namely, those dipoles that get excited by the resonant drive (whose frequency can be swept on demand) can become irreversibly trapped in dark states (at sufficiently low temperatures).
The difference concerning the newly proposed STIRAP-based SHB \cite{molmer2019}, or the single-photon pulse that we have studied here, is the quantum coherent nature of the process.
Understanding how STIRAP-based SHB depends on temperature seems a valuable opportunity for widening the concept of quantum dissipative adaptation (similarly, Ropp {\it et al.} \cite{naturephoton2018} show a groundbreaking experiment on temperature-dependent dissipative self-organization in optical space).

Optomechanical nanoresonators \cite{nnanoJMG} also hold the promise of displaying some kind of quantum dissipative adaptation in the lines of our results.
We glean this notion from Kedia {\it et al.} \cite{bistableDA}, where signatures of dissipative adaptation are shown in disordered networks of classical bistable springs.
In the mechanical nanoresonators by Yeo {\it et al.} \cite{nnanoJMG}, bistability can arise from a strain-mediated coupling between the centre of mass of an oscillating nanowire and the quantum state of a single semiconductor quantum dot embedded therein.
This kind of coupling between an optically controllable microscopic degree of freedom (within the quantum dot) and a mesoscopic degree of freedom (the nanoresonator centre of mass) opens appealing perspectives for studying dissipative adaptation at a quantum-classical boundary of a many-body system.
As a last word on resonators, it seems relevant investigating whether the vibration-assisted exciton transport found in photosynthetic light-harvesting antennae \cite{AOC14} could be related with a quantum dissipative adaptation.

To conclude, the quantum dissipative adaptation we have found can be regarded as a proof of principle, in need of generalizations towards many directions.
As a first example, it would be worth investigating dissipative adaptation in larger Hilbert spaces (i.e., in a vast energy landscape, in the words of \cite{naturenano2015, PRL2017}).
Multistability makes self-organization in classical many-body systems a fascinating problem \cite{naturenano2015,PRL2017, bistableDA}.
Curiously, the multistability from complex classical systems is reminiscent of our zero-temperature model.
Among the huge amount of available ``non-organized'' stable states that our $\Lambda$ system can occupy, only a single (exceptional) state is populated when the work source is optimal.
In other words, although we find an infinite number of (pure or mixed) combinations of states $\ket{a}$ and $\ket{b}$ that are equally stationary in the non-driven zero-temperature environment, the system always ends up in the rare state $\ket{a}$ (irrespective of the infinitely many possible initial states) once it has been driven by the suitable pulse (which maximizes the absorbed work in the $\ket{a}$ to $\ket{b}$ transition).

It also remains to be investigated how other sources of work and finite temperatures affect dissipative adaptation in quantum systems.
In our $\Lambda$ system, for instance, we expect that the forward ($\ket{a}$ to $\ket{b}$) transition will keep being favored with respect to the backward ($\ket{b}$ to $\ket{a}$) transition at finite temperatures, as long as the work source is strong enough.
Also, the time dependence of this asymmetry should be transient or stationary, depending on the work source type (pulsed or continuous).
In quantum thermodynamics, the influence of the environment temperature on entropy production at the quantum regime, on the one hand, has been the subject of many recent studies \cite{deffner11,goold15,santos19,landi20,auffeves20}.
The focus on model-independent aspects is justified not only due to the interest in generality, but also due to the closest possible analogies with classical fluctuation theorems and with the Landauer erasure principle (that establishes how information erasure is connected to thermodynamic entropy production due to heat dissipation).
On the other hand, Manzano {\it et al.} \cite{parrondo18}, for instance, provide an alternative formalism for quantum fluctuation theorems that go beyond thermal-equilibrium states of the environment.
It may be the case that the quantum fluctuation theorems developed in the papers above (and in the references therein) present fruitful methods in the search for a general quantum theory of dissipative adaptation.
In summary, we provide the starting point towards a quantum thermodynamics of driven self-organization.

\section{Methods}

{\bf Global system-plus-reservoir quantum dynamics.}
To obtain explicit expressions for the transition probabilities in Eqs.(\ref{pae}), we need to solve the Schr\"odinger equation 
$i\hbar \partial_t \ket{\xi(t)} = H\ket{\xi(t)}$, 
where
$\ket{\xi(t)} = \psi(t) \ket{e,0} 
+ \sum_\omega \phi_\omega^a(t) a_\omega^\dagger \ket{a,0}
+ \phi_\omega^b(t) b_\omega^\dagger \ket{b,0}$.
After a Wigner-Weisskopf approximation and using that
$\phi_n(z,t) \equiv \sum_\omega \phi_\omega^n(t) \exp(ik_\omega z)$, with a linear dispersion relation, 
$\omega = c k_\omega$, and 
$\sum_\omega \ra \int d\omega \varrho_\omega \approx \varrho \int d\omega$, we find that

\beq
\partial_t \psi(t) =  - \left( \frac{\Gamma_a + \Gamma_b}{2} + i\omega_a \right) \psi(t) - g_a \phi_a(-ct,0),
\label{eqpsi}
\eeq
with

\begin{align}
\phi_a(z,t) &= \phi_a(z-ct,0) \nn \\
&+ \sqrt{2\pi \varrho \Gamma_a} \Theta(z) \Theta(t-z/c) \psi(t-z/c)
\label{phia}
\end{align}
and

\begin{align}
\phi_b(z,t) &= \phi_b(z-ct,0) e^{-i\delta_{ab} t} \nn \\
&+ \sqrt{2\pi \varrho \Gamma_b} \Theta(z) \Theta(t-z/c) \psi(t-z/c) e^{-i\delta_{ab} z/c},
\label{phib}
\end{align}
where $\Gamma_{k} = 2\pi g_k^2 \varrho$.

Integrating Eq.(\ref{eqpsi}) for $\psi(0)=0$ gives 

\beq
\psi(t) = - g_a \int_0^t \phi_a(-ct',0) e^{-\left( \frac{\Gamma_a+\Gamma_b}{2} + i\omega_a \right)(t-t')} dt',
\label{psimethods}
\eeq
which depends on the initial photon wavepacket,

\beq
\phi_a(z,0) = \phi^{\mathrm{shape}}_a(z,0) e^{i\omega_L \frac{z}{c}}.
\label{shape}
\eeq
$\omega_L$ is the central frequency of the pulse and $ \phi^{\mathrm{shape}}_a(z,0)$ is its spatial shape.
For the sake of providing an explicit and physically motivated example, we consider in Eq.(\ref{pabtinfty}) the shape of the initial photon wavepacket to be an exponential (raising in space, equivalent to decaying in time for a right-propagating pulse), as typical from spontaneous emission,

\beq
\phi^{\mathrm{shape}}_a(z,0)  = N \Theta(-z) e^{\frac{\Delta}{2} \frac{z}{c}},
\label{exp}
\eeq
where $\Delta$ is the pulse linewidth and 
$N = \sqrt{2\pi \varrho \Delta}$ is a normalization factor (considering $\phi_b(z,0) = 0$).
Finally, and substituting $\Gamma_{a} = 2\pi g_a^2 \varrho$, we have that

\beq
\psi(t) = 
- f_\Delta \
e^{-\left( \frac{\Gamma_a + \Gamma_b}{2} + i \omega_a \right) t} 
\left[ e^{\left( \frac{\Gamma_a+\Gamma_b-\Delta}{2} - i\delta_L \right)t} - 1 \right],
\eeq
where

\beq
f_\Delta \equiv \frac{\sqrt{\Gamma_a \Delta}}{\frac{\Gamma_a+\Gamma_b - \Delta}{2} - i \delta_L}
\eeq
and $\delta_L \equiv \omega_L - \omega_a$.

The average interaction energy is given by
\beq
\langle H_I(t) \rangle \equiv \mbox{tr}[\rho(t) H_I].
\eeq
From Eqs.(\ref{rhozero}), (\ref{rhoglobal}), and the definition $\ket{\xi(t)} \equiv U\ket{a,1_a}$, we have that
\beq
\langle H_I(t) \rangle = p_a^{(0)} \bra{\xi(t)} H_I \ket{\xi(t)} + p_b^{(0)} \bra{b, 1_a} H_I \ket{b, 1_a}.
\eeq
Using $H_I$ from Eq.(\ref{HI}), we have that $\bra{b, 1_a} H_I \ket{b, 1_a} = 0$ and $\bra{\xi(t)} H_I \ket{\xi(t)} = 2 (\hbar g_a \mbox{Im}[\psi^*(t) \phi_a(-ct,0)]$ $+$ $\hbar g_b \mbox{Im}[\psi^*(t) \phi_b(-ct,0)])$.
Choosing $\phi_b(z,0) = 0$ obviously makes $\mbox{Im}[\psi^*(t) \phi_b(-ct,0)] = 0$.
By choosing the resonance condition, $\delta_L = 0$, we have from Eqs. (\ref{psimethods}) and (\ref{shape}) that $\mbox{Im}[\psi^*(t) \phi_a(-ct,0)]= 0$.
This shows that $\langle H_I(t) \rangle = 0$ at resonance.

\

{\bf Entropy of the environment.}
The quantum state of the environment in our model is obtained from the global initial state 
$
\rho(0) = (p_a^{(0)} \ket{a}\bra{a} + p_b^{(0)} \ket{b}\bra{b}) \otimes \ket{1_a} \bra{1_a}
$.
We obtain that

\begin{align}
\rho_E(t) &= p_a^{(0)} |\psi(t)|^2 \ket{0}\bra{0} + p_a^{(0)} N_b(t) \ket{\tilde{1}_b} \bra{\tilde{1}_b} \nn \\
&+ p_a^{(0)} N_a(t)  \ket{\tilde{1}_a} \bra{\tilde{1}_a} + p_b^{(0)} \ket{1_a^{\mathrm{free}}} \bra{1_a^{\mathrm{free}}}.
\label{rhoE}
\end{align}
We have defined
$
N_k(t) \equiv \sum_\omega |\phi_\omega^k(t)|^2
$, with

\beq
\ket{\tilde{1}_k} \equiv N_k^{-1/2} \sum_\omega \phi_\omega^k (t) k^\dagger_\omega \ket{0},
\label{tilde1k}
\eeq
and

\beq
\ket{1_a^{\mathrm{free}}} \equiv  \sum_\omega \phi_\omega^a(0) \exp(-i\omega t)\ a^\dagger_\omega \ket{0}.
\label{1afree}
\eeq

To explicitly calculate the von Neumann entropy, 

\begin{align}
S_E &= - \mbox{tr}[\rho_E(t) \ln \rho_E(t)], \nn\\
&= -\sum_j \lambda_j \ln \lambda_j,
\label{vNlambdas}
\end{align} 
we need to find the eigenvalues $\lambda_j$ of $\rho_E$.
The exact diagonalization of $\rho_E$ gives us four non-zero eigenvalues, namely, 
$\lambda_1 =  p_a^{(0)} |\psi(t)|^2$,
$\lambda_2 = p_a^{(0)} N_b(t)$,
and

\begin{align}
\lambda_{3,4} &= \frac{p_a^{(0)} N_a + p_b^{(0)}}{2} \nn\\
&\pm \frac{1}{2} \sqrt{\left( p_a^{(0)} N_a-p_b^{(0)} \right)^2 + 4 p_a^{(0)} p_b^{(0)} N_a |\bra{1^{\mathrm{free}}_a} \tilde{1}_a \rangle|^2}.
\label{lambda34}
\end{align}
Terms $N_a$ and $N_b$ can be interpreted as the average number of photons in each continuum of modes.
Their mathematical expressions, however, are invariably connected with the three-level system probabilities, namely, 

\beq
N_k(t) = p_{a\ra k}(t).
\eeq
We calculate the overlap between the states representing the free propagation and the reemitted photon at mode $a$ in the real-space representation,

\beq
\bra{1^{\mathrm{free}}_a} \tilde{1}_a \rangle = 
\frac{1}{\sqrt{N_a}} \frac{1}{2 \pi \varrho c} \int_{-\infty}^\infty dz \ \phi^*_{\mathrm{a, free}}(z,t) \phi_a(z,t),
\eeq
where $\phi_{\mathrm{a,free}}(z,t) \equiv \phi_a(z-ct,0)$.

With the aid of (\ref{eqpsi}) and (\ref{phia}), and using integration by parts, we find that

\beq
\bra{1^{\mathrm{free}}_a} \tilde{1}_a \rangle {\sqrt{N_a}} = 
1 - \left(\frac{\Gamma_a+\Gamma_b}{2\Gamma_b}\right) p_{a\ra b}(\infty),
\label{overlap}
\eeq
valid at long times $t\ra \infty$.
Eq.(\ref{overlap}) provides the core connection between the overlap and the transition probability at long times that we needed.
Additionally, $N_a(\infty) = 1 - N_b(\infty) = 1 - p_{a\ra b}(\infty)$ and $\lambda_1(\infty) = 0$.
Since Eqs. (\ref{adaptationwork}) and (\ref{ec2}) link $p_{a\ra b}(\infty)$ to $\langle {Q}_{\mathrm{diss}}\rangle_{a}$, 

\beq
p_{a\ra b}(\infty) = \left( \hbar\omega_a \frac{\Gamma_a+\Gamma_b}{\Gamma_b}  - \hbar\delta_{ab} \right)^{-1} \langle {Q}_{\mathrm{diss}}\rangle_{a},
\label{pQgen}
\eeq
we have thus analytically established the function
$S_E(\infty)$ vs. $\langle {Q}_{\mathrm{diss}}\rangle_{a}$,
valid at long times $t\ra \infty$, under the resonant condition $\omega_L = \omega_a$, for any arbitrary pulse shape.

Finally, the classical contribution to the entropy is
$
S_E^c \equiv S_E - p_a^{(0)} S(\mbox{tr}_S[\ket{\xi(t)} \bra{\xi(t)}])
$,
where

\begin{align}
&S(\mbox{tr}_S[\ket{\xi(t)} \bra{\xi(t)}])  \nn\\
&= -\left( N_a \ln N_a + N_b \ln N_b + |\psi(t)|^2 \ln |\psi(t)|^2 \right),
\label{qvN}
\end{align}
for which we also have the analytic solution.

\section{Data availability}
Data sharing not applicable to this article as no datasets were generated or analysed during the current study.

\section{Code availability}
The code used to produce the figure in this article is available from the corresponding author upon request.

\

\section{Author contributions}
D. V., F. B. and T. W. contributed to all aspects of the research, with the leading input of D. V.

\section{Competing interests}
The authors declare no competing interests.

\section{acknowledgements}
This work was supported by the Serrapilheira Institute (Grant No. Serra-1912-32056) and by Instituto Nacional de Ci\^encia e Tecnologia de Informa\c c\~ao Qu\^antica (INCT-IQ), Brazil.


%

%

%

%

%
\end{document}